\documentclass[twocolumn,amsmath,amssymb,aps, prl,
showkeys,showpacs,nofootinbib] {revtex4-1}
\usepackage{graphicx}
\usepackage{bm}
\usepackage[colorlinks=true,breaklinks=true,allcolors=blue]{hyperref}
\usepackage{dcolumn}
\begin{document}
\title{Quantum phase transitions of the Majorana toric code in the presence of finite
Cooper-pair tunneling}
\author{Ananda Roy}
\email{roy@physik.rwth-aachen.de}
\affiliation{JARA Institute for Quantum Information, RWTH Aachen University, 52056 Aachen, Germany}

\author{Barbara M. Terhal}
\affiliation{JARA Institute for Quantum Information, RWTH Aachen University, 52056 Aachen, Germany}
\author{Fabian Hassler}
\affiliation{JARA Institute for Quantum Information, RWTH Aachen University, 52056 Aachen, Germany}
\begin{abstract}
The toric code based on Majorana fermions on mesoscopic superconducting islands is
a promising candidate for quantum information processing. In the limit of
vanishing Cooper-pair tunneling,  it has been argued that the phase
transition separating  the topologically ordered phase of the toric code from
the trivial one is in the universality class of (2+1)D-XY. On the other hand,
in the limit of infinitely large Cooper-pair tunneling, the phase transition
is in the universality class of (2+1)D-Ising. In this work, we treat the case
of finite Cooper-pair tunneling and address the question of how the continuous
XY symmetry breaking
phase transition turns into a discrete $\mathbb{Z}_2$ symmetry breaking one
when the Cooper-pair tunneling rate is increased.  We show that this happens
through a couple of tricritical points and first order phase transitions.
Using a Jordan-Wigner transformation, we map the problem to that of spins
coupled to quantum rotors and subsequently, propose a Landau field theory for
this model that matches the known results in the respective limits.  We
calculate the effective field theories and provide the relevant critical
exponents for the different phase transitions. Our results are relevant for
predicting the stability of the topological phase in  realistic experimental
implementations.
\end{abstract}
\maketitle 

\begin{figure}
  \includegraphics[width = .95\linewidth]{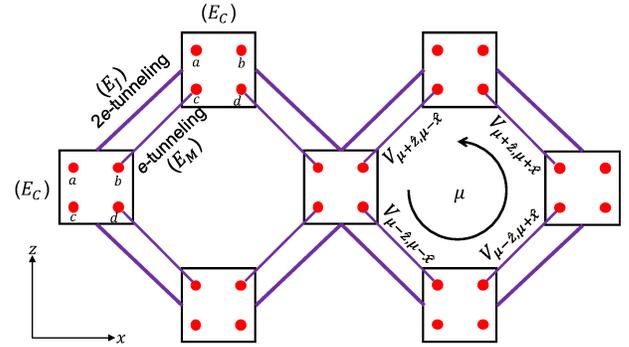}
\caption{\label{fig_1} (Color online) Two plaquettes of an infinite lattice of
Majorana fermions (denoted by red dots) on superconducting islands (denoted by
white squares) are shown. The charging energy of each island is denoted by
$E_C$. The thin links between the Majoranas
denote single-electron tunneling at a rate $E_M$ and the
thick links between two neighboring islands indicate Cooper-pair tunneling
at a rate $E_J$. The four Majorana interaction terms around a plaquette are
indicated [see Eq.~\eqref{vij}].  }
\end{figure}

The toric code  \cite{Kitaev2003, Kitaev2006} is a promising candidate for
fault-tolerant quantum computation \cite{DiVincenzo2009, Fowler_Cleland_2012}.
It describes a topologically ordered system whose four-fold degenerate ground
state is protected from local perturbations.  In contrast to standard
realizations of the toric code using qubits, an alternative approach has been
proposed using interacting Majorana fermions \cite{Xu2010, Terhal2012,
Vijay2015, Landau2016, Karzig2016a, Litinski2017}.  This approach considers
a 2D array of Majorana fermions on mesoscopic superconducting islands (see
Fig.~\ref{fig_1}). Each island has a charging energy $E_C = e^2/2C$, where $C$
is the capacitance of each island to a ground plane. The Majorana fermions
enable tunneling of single electrons between two neighboring islands
\cite{Fu2010}, in addition to Cooper-pair tunneling. The rates of
single-electron and Cooper-pair tunneling are denoted by $E_M$ and $E_J$,
respectively. In the limit of vanishing Cooper-pair tunneling rate and large
charging energy, the system supports a topologically ordered phase described
by an effective toric code Hamiltonian \cite{Xu2010, Landau2016}. Furthermore, in
this limit, upon increasing the single-electron tunneling rate, the system goes through
a zero-temperature, topological phase transition in the universality class of (2+1)D-XY
\cite{Xu2010}. On the other hand, in the limit of infinite Cooper-pair
tunneling rate, the system also shows the topologically ordered toric
code phase and the transition to the trivial phase is in the universality
class of (2+1)D-Ising \cite{Terhal2012}.  In this work, we address the
question how the topological phase transition that breaks continuous XY 
symmetry gets transformed into one that breaks discrete
$\mathbb{Z}_2$ symmetry when the Cooper-pair tunneling rate is increased. This question is, in
particular, nontrivial as the system undergoes a  Mott
insulator-superconductor transition when increasing the Cooper-pair tunneling rate with respect to
the charging energy \cite{Fisher1989, Fazio1997, Fazio2001, Herbut2007, Sachdev2011}.
Starting from the microscopic Hamiltonian, we present a symmetry-based,
phenomenological, Ginzburg-Landau field-theoretic description
that captures the critical behavior of the system in the presence of both
single-electron and Cooper-pair tunneling.  We show that the XY phase
transition transforms to an Ising phase transition through a couple of
tricritical points and first order phase transitions.  

The paper is outlined as follows. From the microscopic Hamiltonian of the
system, after a Jordan-Wigner transformation, we map the problem to
coupled spins and rotors, with nearest-neighbor interactions. From symmetry
considerations, we propose a field theory that describes the critical behavior
of the system. Subsequently, we derive effective field theories for the
different phase transitions. Finally, we discuss experimental signatures of the phase transitions. Throughout this work,
we restrict ourselves to zero temperatures.

The Hamiltonian of the system is given by $H = H_C + H_J + H_M$, where 
\begin{align}\label{h0} 
  H_C &= 4E_C\sum_i n_i^2,\qquad H_J = -E_J\sum_{\langle
i,j\rangle} \cos(\phi_i-\phi_j), \nonumber\\ 
H_M &= -E_M \sum_{\langle
i,j\rangle}V_{ij}\cos\Big(\frac{\phi_i-\phi_j}{2}\Big).  
\end{align} 
Here, the superconducting phase $\phi_i$ and the excess charge $n_i$ (in units
of Cooper pairs) on the $i$-th island are canonically conjugate. In this work,
we treat the idealized case of zero offset charges in the  absence of
disorder. The Majorana tunneling operator $V_{ij}$ between the two neighboring
islands $i,j$ is given by (see Fig. \ref{fig_1}) \cite{Terhal2012} 
\begin{align} \label{vij}
V_{\mu+\hat{z}, \mu-\hat{x}} &= i
\gamma_b^{\mu-\hat{x}}\gamma_c^{\mu+\hat{z}},& V_{\mu-\hat{z},\mu-\hat{x}} &=
i\gamma_a^{\mu-\hat{z}}\gamma_d^{\mu-\hat{x}},\nonumber\\ V_{\mu-\hat{z},
\mu+\hat{x}} &= i \gamma_c^{\mu+\hat{x}}\gamma_b^{\mu-\hat{z}},&
V_{\mu+\hat{z},\mu+\hat{x}} &= i\gamma_d^{\mu+\hat{z}}\gamma_a^{\mu+\hat{x}},
\end{align} 
where the $\gamma^i_\alpha$ are Hermitian, Majorana fermion
operators obeying $\{\gamma^i_\alpha, \gamma^j_\beta\} = \delta_{ij}
\delta_{\alpha\beta}$. The fermion parity on the $i$-th island is given by the
operator  $\mathcal{P}_i = -\gamma^{i}_a\gamma^{i}_b\gamma^{i}_c\gamma^{i}_d$.
As the charge is constraint by the fermion parity, the (physical) Hilbert-space
for the Hamiltonian $H$ is spanned by the wavefunctions satisfying
$\psi(\phi_i+2\pi) = (-1)^{(1-\mathcal{P}_i)/2}\psi(\phi_i)$ \cite{Fu2010}. At
finite charging energy, the ground state is in the even parity sector on each
island (${\cal P}_i\equiv +1$). In this sector, the four Majorana fermions on
each island encode one qubit \cite{Bravyi2006} and, neglecting $H_J$, a
perturbation calculation in $E_M/E_C$ yields the toric code Hamiltonian
\cite{Xu2010, Landau2016}.

Going beyond this perturbation analysis, first, we perform  a gauge transformation in order to simplify
the Hilbert-space to $2\pi$-periodic functions \cite{VanHeck2012}. Then, we map the Majorana fermions to spins using a 
Jordan-Wigner transformation  (for
details, see Supplement of \cite{Terhal2012}). As a result, we arrive at the
Hamiltonian
\begin{align}
\label{hc}
H_C &= 4E_C\sum_i \Big(n_i+\frac{1+\sigma_i^z}{4}\Big)^2,\\
H_M &= -\frac{E_M}{2} \sum_{\langle
i,j\rangle}s_{i,j}\Big\{\sigma_i^-\sigma_j^-(e^{i\phi_i} +
e^{i\phi_j})\nonumber\\& \qquad\qquad\qquad
+\sigma_i^-\sigma_j^+\big[1+e^{i(\phi_i-\phi_j)}\big]
+ \text{H.c.}\Big\}, \nonumber
\end{align}
while $H_J$ remains invariant \cite{Cooper_pair_note}.  Here, the sign of the interaction is
determined by gauge bits $s_{i,j}=\pm1$ \cite{Terhal_note}.  This form of the
Hamiltonian is most useful for numerical analysis and makes the symmetries of
the model explicit. Most importantly, the Hamiltonian is invariant under the
simultaneous transformations $U_\theta\colon e^{i\phi_i}\mapsto
e^{i\phi_i}e^{i\theta}$, $\sigma_i^+\mapsto \sigma_i^+e^{i\theta/2}$.
Physically, this global symmetry originates from the fact that the spins
correspond to  single-electrons that carry half of the charge of the Cooper
pairs. 

The three terms in the Hamiltonian give rise to phases which can be classified
according to how they break the $U_\theta$ symmetry. In the phase where
the Cooper-pair tunneling $H_J$ aligns the rotors $e^{i\phi_i}$, the
$U_\theta$ symmetry is spontaneously broken. The ground state is only
invariant under $U_\theta$ with $\theta$ being a multiple of $2\pi$. We  denote this
phase by $\{2\pi\}$. The single-electron tunneling $H_M$, on the other hand,
orders the spins, with $\sigma^+_i$ obtaining a finite expectation value, such
that the ground state is only invariant with $\theta$ being a multiple of $4\pi$. We
denote this phase by $\{4\pi\}$. The Coulomb interaction $H_C$ disorders both
the rotors and the spins and restores the full symmetry of the ground state
under $U_\theta$ for all $\theta$. We denote this phase by
$\{\theta\}$.

Next, we discuss the signatures of the three phases in the charge sector and the
relation to $\mathbb{Z}_2$ spin liquids. In the phase $\{\theta\}$, the strong
Coulomb interactions localize all charges and turns the system into a Mott
insulator. Going over to the phase $\{2\pi\}$, the charges condense into
Cooper pairs turning the system into a superconductor of charge $2e$.  In the
phase $\{4\pi\}$ where the spins are ordered, the condensate is comprised of
charge $e$-bosons (also called `holons' or `chargons')
\cite{Wen1991,Balents1999, Senthil2000}.  In the language of $\mathbb{Z}_2$
spin liquids, in our model,  the `vison' excitation on the plaquette $\mu$,
given by $\prod_{\square_\mu} s_{i,j}=-1$, is static. Physically, the vison
indicates the presence of a superconducting vortex in the plaquette. The
phases $\{\theta\}, \{2\pi\}$ correspond to the deconfined phase where visons
can be separated from each other without energy cost. On the other hand, in
the confined phase $\{4\pi\}$,  the energy associated with two vison
excitations increases with the spatial separation between them
\cite{Fradkin1979}. The system is in the toric code phase when the
visons are deconfined \cite{Read1991, Wen1991} or equivalently, when the spin
sector is ordered \cite{Terhal2012}.  In what follows, we derive an effective Landau field theory for
the model. We drop explicit reference to vison degrees of freedom and infer topological ordering from the ordering in the spin sector.

\begin{figure}
  \includegraphics[width = .95\linewidth]{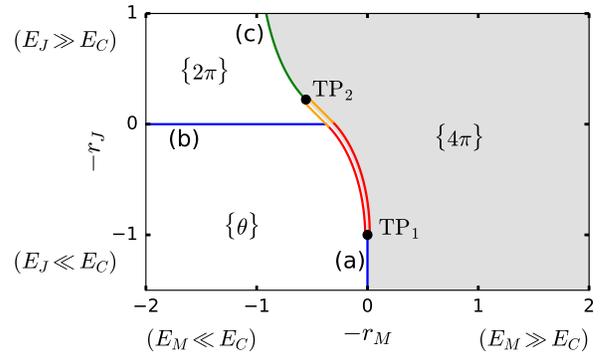}
\caption{\label{fig_2} (Color online) Phase diagram for the model as a
function of $r_J,r_M$ for $u_\text{s}=u_\text{r}=\alpha=\beta=1$. The blue
line, marked with (a), denotes a (2+1)D-XY phase transition line, separating
the phases $\{\theta\}$ and $\{4\pi\}$. This transition terminates
in a tricritical point ($\rm{TP_1}$), after which the transition becomes a
first-order transition (shown as red double line).  The blue line, marked with
(b), also denotes a (2+1)D-XY phase transition line, that separates the phases
$\{\theta\}$ and $\{2\pi\}$. This transition line terminates at the
first-order line coming out of $\rm{TP_1}$. The green line, marked with (c),
denotes a (2+1)D-Ising transition line separating the phases $\{2\pi\}$ and
$\{4\pi\}$. This phase transition line terminates in a tricritical point
($\rm{TP_2}$), after which turns into a first-order line (shown as orange
double line), which smoothly transforms into the other first-order line. As
discussed in the main text, both $\{\theta\}$ and $\{2\pi\}$ are topologically
ordered. A similar
phase-diagram was proposed in \cite{Fradkin_note}.  }
\end{figure}

We consider complex fields 
$\psi_\text{r}(\bm{r},\tau)$ and $\psi_\text{s}(\bm{r},\tau)$ which correspond
to the coarse-grained expectation values of $e^{i\phi_i}$ and $\sigma^+_i$ in
imaginary time $\tau$ respectively. Interested in the behavior of the system close to the point where all
the three phases meet, we expect the relevant degrees of freedom to be given
by the low-frequency, long-wavelength behavior of these complex fields. 
The microscopic symmetry $U_\theta$ is
elevated to the symmetry $\psi_\text{r}\mapsto \psi_\text{r} e^{i\theta},
\psi_\text{s}\mapsto \psi_\text{s} e^{i\theta/2} $ on the coarse-grained
variable that has to be respected in the effective field theory. Close to
the phase transition, the fields are small. Thus, we perform a Taylor and gradient
expansion in $\psi_\text{r}, \psi_\text{s}$.
 The partition function at zero temperature is given by $Z = \int\! {\cal
D}\psi_\text{s}\,{\cal D}\psi_\text{s}^*\,{\cal D}\psi_\text{r}\,{\cal
D}\psi_\text{r}^*\,e^{-S}$ with the Euclidean action
\begin{align}\label{action_tot}
  S= & \int\! d^2r\, d\tau \Bigl[  
    |\partial_\tau\psi_\text{s}|^2 + |\partial_\tau\psi_\text{r}|^2  
    + K_\text{s}|\nabla\psi_\text{s}|^2+K_\text{r}|\nabla\psi_\text{r}|^2
    \nonumber \\
   &+r_M |\psi_\text{s}|^2 + 
   r_J |\psi_\text{r}|^2  + u_\text{s}|\psi_\text{s}|^4+
   u_\text{r}|\psi_\text{r}|^4 
   +\beta|\psi_\text{s}|^2|\psi_\text{r}|^2 \nonumber\\
   &- \alpha({\psi_\text{s}^*}^2\psi_\text{r} +
 \psi_\text{s}^2\psi_\text{r}^*)  \Bigr].
\end{align}
We assert that the terms in Eq.~\eqref{action_tot} exhaust the relevant terms
up to quartic order in the fields, consistent with the symmetries of the
Hamiltonian, that can appear in the action. Note that only modulus-square of the first-order imaginary-time derivatives of the fields 
appear in the action \cite{Xu2010, 2timeder}. The first two lines of Eq.~\eqref{action_tot} is the theory of
the tetracritical point (see Chap. 4 of \cite{Chaikin2000}). The cubic term in
the third line has, to our knowledge, not been investigated and is crucial for
the prediction of the phase-diagram of the system.  In order to have a stable
theory, $u_\text{s}, u_\text{r}, \beta$ must be positive and we choose
$\alpha>0$ without loss of generality. The parameter $r_x$ is used to tune
through the phase transition and corresponds to $-E_x/E_C$, where $x = M,J$.
The phase diagram of the model is given in Fig.~\ref{fig_2} and the phase
transitions will be analyzed below.

Before deriving effective field theories for the different phase transitions,
we qualitatively explain how the field theory [Eq. \eqref{action_tot}]
describes the different phases of the system. For positive $r_J$ and $r_M$,
the fields $\psi_{\rm s},\psi_{\rm r}$ vanish and the system is in the
$\{\theta\}$ phase. When $r_M$ changes sign [across the line (a) in
Fig.~\ref{fig_2}], the spin orders and $|\psi_s|$ attains a finite value
resulting in the phase $\{4\pi\}$. In this transition, $\psi_\text{r}$ is
slaved to $\psi_\text{s}$ since, due to the cubic term, $\psi_\text{s}^*{}^2$
acts as a `magnetic field' for $\psi_\text{r}$. Note that the phases
$\theta_\text{s}, \theta_\text{r}$ of the complex order parameters
$\psi_\text{s}, \psi_\text{r}$ are locked via $ \theta_\text{r} = 2
\theta_\text{s}$.  On the other hand, when $r_J$ changes sign [across the line
(b) in Fig.~\ref{fig_2}] for $r_M$ large enough, only the field
$|\psi_\text{r}|$ becomes finite and the system is in the phase $\{2\pi\}$.
Lowering $r_M$ reduces the stability of the disordered phase in the spin
sector until the spin orders. The system enters the phase $\{4\pi\}$ via the
line (c) in Fig.~\ref{fig_2} with the spin and the rotor order parameters
phase locked as described above. In the following, we derive effective field
theories for each of the transitions. This will provide information about the
nature of the phase-transitions and the various associated critical exponents.

First, we analyze the phase transition between the phases $\{\theta\}$ and
$\{4\pi\}$.  To get an effective theory for $\psi_\text{s}$, we integrate out
the rotor field $\psi_\text{r}$ by considering small fluctuations
$\psi_\text{r} = \bar{\psi}_\text{r} + \delta\psi_\text{r}$ around the saddle
point $\bar{\psi}_\text{r}$ with $\delta S=0$. In the vicinity of the phase
transition, to leading order, the saddle point solution is given by
$\bar{\psi}_\text{r}\propto {\psi_\text{s}}^2$.  Substituting
$\bar{\psi}_\text{r}$ in Eq.~\eqref{action_tot} and integrating out the
Gaussian fluctuations $\delta\psi_\text{r}$ keeping the lowest order terms in
$\alpha$, $\beta$ in the cumulant expansion, the partition function assumes
the form $Z^{(a)} = \int\! {\cal D}\,\psi_\text{s}\,{\cal D}\psi_\text{s}^*\,
e^{-S^{(a)}}$, where the action is given by
\begin{multline}
%\label{sa}
S^{(a)}=\int\! d^2r\,
d\tau\Big\{|\partial_\tau\psi_\text{s}|^2+K_\text{s}|\nabla\psi_\text{s}|^2 +
r_M |\psi_\text{s}|^2\nonumber\\
+ \Big(u_\text{s}-\frac{\alpha^2}{r_J}\Big)|\psi_\text{s}|^4 +
\frac{\alpha^2\beta}{r_J^2}|\psi_\text{s}|^6 \Big\}\nonumber.
\end{multline}
As long as the prefactor of the quartic term is positive, \textit{i.e.}, for
$r_J>\alpha^2/u_\text{s}$, the sextic term is irrelevant and the phase
transition at $r_M=0$ is a second order (2+1)D-XY transition.  The phase
transition line [marked by (a) in Fig.~\ref{fig_2}] terminates at the
tricritical point ($\rm{TP}_1$) given by $r_J = \alpha^2/u_\text{s}$. Lowering
$r_J$ further, the quartic term changes sign and the transition becomes first
order (see Chap. 4 of \cite{Chaikin2000}).
 
Next, we analyze the phase transition between the phases $\{\theta\}$ and
$\{2\pi\}$ [marked by (b)]. Across this transition, $\psi_\text{s}$ stays zero,
while $|\psi_\text{r}|$ turns finite. As before, integrating over small
fluctuations $\delta\psi_\text{s}$ around the saddle point $\bar{\psi}_\text{s} = 0$, we get an
effective partition function $Z^{(b)} = \int\! {\cal D}\psi_\text{r}\,{\cal
D}\psi_\text{r}^*\, e^{-S^{(b)}}$ with
\begin{align}
%\label{sb}
S^{(b)} &= \int\! d^2r\, d\tau\big(|\partial_\tau\psi_\text{r}|^2+K_\text{r}|\nabla\psi_\text{r}|^2
+ r_J |\psi_\text{r}|^2 + u_\text{r}|\psi_\text{r}|^4\big)\nonumber.
\end{align}
Thus, this transition is the Bose-Hubbard phase transition \cite{Fisher1989, Herbut2007,
Sachdev2011}, and the phase transition line is given by $r_J=0$.  This phase
transition line terminates at the first order line coming out of $\rm{TP_1}$
\cite{tp3}.

Now, we analyze the phase transition [marked by (c)] between the phases
$\{2\pi\}$ and $\{4\pi\}$. To get the effective field theory, we use the
parametrization $\psi_\text{r} = (\bar{\rho}_\text{r} + \delta
\rho_r)e^{i\theta_\text{r}/\bar{\rho}_\text{r}}$, $\psi_\text{s} =
(\sigma+iw)e^{i\theta_\text{r}/2\bar{\rho}_\text{r}}$,  where
$\bar{\rho}_\text{r}$ is the saddle point value of $|\psi_\text{r}|$ and the
real fields $\delta \rho_\text{r}$, $\theta_\text{r}$ and $\sigma$, $w$ denote
the fluctuations of $\delta\psi_\text{r}$ and $\delta\psi_\text{s}$.  The
fluctuations in $\theta_{\rm{r}}$ correspond to the massless Goldstone mode
associated with the symmetry breaking in the rotor sector. They decouple from
the rest.  Integrating over $v,w$, we arrive at the partition function
$Z^{(c)}=\int\!{\cal D}\theta_\text{r}\,
e^{-\int\!d^2r\,d\tau(|\partial_\tau\theta_\text{r}|^2+K_\text{r}|\nabla\theta_\text{r}|^2)/2}\int\!{\cal
D}\sigma\, e^{-S^{(c)}}$ with
\begin{align}
S^{(c)} \!&=\!\!  \int\! d^2r\,
d\tau\big\{(\partial_\tau\sigma)^2\!+\!K_\text{s}(\nabla\sigma)^2 +t_c \sigma^2
  \!+\! u_c\sigma^4 \!+\!\tilde{u}_c\sigma^6
 \big\}\nonumber. 
\end{align}
We see that the phase transition is described by the emergent Ising degree of
freedom $\sigma$. In particular, the field $\sigma$ acquires a finite value
when $t_c$ changes sign. The Ising degree of freedom corresponds to the two
possibilities $\theta_s =\frac12  \theta_r$ and $\theta_s =\frac12
\theta_r+\pi$ of phase-locking of the spin order parameter with the rotor
field. To lowest order, the parameters in the action $S^{(c)}$ are related to
the parameters of the original field via
\begin{align}
 t_c &= r_M-\alpha\sqrt{\frac{-2r_J}{u_\text{r}}}-\frac{\beta
 r_J}{2u_\text{r}}, \qquad \tilde u_c = \frac{u_\text{s}(2\sqrt{u_\text{s}
 u_\text{r}}+\beta)^3}{4 \alpha^2 u_\text{r}},\nonumber\\ 
 u_c &=
 u_\text{s}+\frac{\alpha^2}{2r_J}-\frac{\alpha\beta}{\sqrt{-2u_\text{r}r_J}}-\frac{\beta^2}{4u_\text{r}}.
\end{align}
For $u_c>0$, \textit{i.e.}, $r_J < r^*_J = -2u_\text{r}\alpha^2/(2\sqrt{u_\text{s}
u_\text{r}}+\beta)^2$, the quartic term is positive and the phase transition
at $t_c=0$ is of second order (2+1)D-Ising type. The phase transition line (c)
terminates at a tricritical point ($\rm{TP_2}$) when $u_c=0$ after which,
\text{i.e.}, for $r_J>r_J^*$, the phase transition turns first order (see
Chap.~4 of \cite{Chaikin2000}).

Finally, we comment on the line of first order phase transition that connects
the two tricritical points TP$_1$ and TP$_2$ in Fig.~\ref{fig_2}. From the
analysis of the field theories $S^{(a)}$ and $S^{(c)}$ close to the
tricritical points, we know that the lines emanate tangential to the second
order lines. Across the first order line separating $\{\theta\}$ and
$\{4\pi\}$, an XY symmetry breaking occurs. This results in both
$\psi_\text{s}, \psi_\text{r}$ being discontinuous. On the other hand, only an
Ising symmetry is broken between the phases $\{2\pi\}$ and $\{4\pi\}$
resulting in only $\psi_\text{s}$ being discontinuous. The nature of the
symmetry breaking changes exactly at the point where the line (b) meets the
first order line. Note that this point is not a tricritical point (see also \cite{tp3}).
 
Before concluding, we comment on experimental signatures of the proposed
phases and phase transitions. As already discussed before, the three phases
$\{\theta\}$, $\{2\pi\}$, and $\{4\pi\}$ correspond to a Mott insulator, a
$2e$-superconductor, and an $e$-superconductor. Thus, they can be
distinguished by their current-voltage characteristic. For instance, the
frequency of the Josephson radiation under a dc voltage bias determines the
charge of the condensate in the superconducting phases while the insulator has
no charge response \cite{Kitaev2001, Kwon2004, Fu2009}. In fact, the Mott-insulator-superconductor transition in Josephson junction arrays in absence of Majorana fermions [transition across (b)] has already been observed \cite{Zant1996}. Measurement of the
superfluid densities given by $|\psi_\text{r}|^2$ and $|\psi_\text{b}|^2$ can
be used to determine the critical exponents $\beta_\text{r}$ and
$\beta_\text{s}$ describing the behavior of the order parameters close to the
phase transition.  The critical exponent $\nu$ that determines the divergence
of the correlation length can be accessed by electromagnetic correlation
measurements as the fields are charged; for example, one can imagine probing
the system by measuring the low-frequency conductance through a pair of
spatially separated capacitive contacts. Close to the phase transition, the
conductance will obtain a finite value for arbitrary distances
\cite{Akhmerov2011}. For convenience, we provide the theoretical values for
the different critical exponents for the three second order phase transitions
and the two tricritical points (using Chap.~5 of \cite{Chaikin2000}) in
Table~\ref{tab1}.
\begin{table}
\begin{ruledtabular}
\begin{tabular}{ccccc}
phase transition&type&$\nu$&$\beta_\text{s}$&$\beta_\text{r}$\\\hline
(a)&(2+1)D-XY&0.67&0.35&0.70\\
(b)&(2+1)D-XY&0.67&--&0.35\\
(c)&(2+1)D-Ising&0.63&0.32&--\\
$\rm{TP_1}$&XY&0.50&0.25&0.5\\
$\rm{TP_2}$&Ising&0.50&0.25&--\\
\end{tabular}
\caption{\label{tab1}Table summarizing the different phase transitions and
tricritical points occurring in the model, their types and the critical
exponents $\nu$ for the correlation length and $\beta_x$, $x = \text{s},
\text{r}$ for the order parameters for the spin and rotor sectors. For
transitions (a) and $\rm{TP_1}$, $\beta_\text{r}=2\beta_\text{s}$, since in the vicinity of the
phase transition, to leading order,
$\bar{\psi}_\text{r}\propto{\psi_\text{s}}^2$.  For the tricritical points,
mean field exponents are exact since (2+1)D is above the upper critical
dimension for the sextic term in $S^{(a)}$ and $S^{(c)}$. }
\end{ruledtabular} 
\end{table}

To summarize, we have analyzed the different phases and the phase transitions
occurring in the Majorana toric code in the presence of Cooper-pair tunneling.
Starting from the microscopic model, we have performed a Jordan-Wigner
transformation and mapped the problem to that of spins coupled to rotors with
nearest neighbor interaction. Subsequently, based on symmetry considerations,
we have proposed a Landau field theory to analyze the critical behavior of the
system at zero temperature. We have shown that as one changes the Cooper-pair
tunneling rate, the topological phase transition separating the toric code phase
from the trivial phase changes from a (2+1)D-XY type to a (2+1)D-Ising type
through a couple of tricritical points and first order transitions. Our
results match the known results in the limiting cases. In particular, we have
provided evidence that the topological order survives for any finite
Cooper-pair tunneling. We have derived an effective field theory for each of
the transitions and commented on the experimental signatures of the phases and
the phase transitions. The present work provides a starting point for further
numerical and field-theoretical investigations of the rich phase diagram of
the Majorana toric code in the presence of Cooper-pair tunneling. Moreover,
with the recent developments in detecting Majorana bound states in solid state
systems \cite{Mourik2012, Albrecht2016}, we are optimistic of experimental
verifications of the field theory predictions.

Discussions with David DiVincenzo, Leonid Pryadko, and Manfred Sigrist are
gratefully acknowledged. AR and BMT acknowledge the support through the ERC
Consolidator Grant No.~682726.


\begin{thebibliography}{10}
\bibitem{Cooper_pair_note}
Note that while $\phi_i$ and $n_i$ remain canonically conjugate, $n_i$ now
  denotes only the excess number of Cooper-pairs on the $i$-th island.
\bibitem{2timeder}
This is because the phase-transition occurs at the tip of the Mott lobe of the
  corresponding Bose-Hubbard model, where the transition is induced by
  increasing the hopping rate $E_J$ \cite{Fisher1989, Herbut2007, Sachdev2011}.
\bibitem{tp3}
The continuous phase transition (b) becomes metastable in the phase $\{4\pi\}$.
  In fact, due to the cubic term, this line of phase transition also terminates
  in a tricritical point after which it turns first order. Depending on
  microscopic parameters, the tricritical point occurs before or after the line
  (b) turns metastable.
\end{thebibliography}
\end{document}